\documentclass{article}
\usepackage[T1]{fontenc}
\usepackage[utf8]{inputenc}
\usepackage{ismir} 
\usepackage{amsmath,cite,url}
\usepackage{graphicx}
\usepackage{color}
\usepackage{etoolbox}
\usepackage{caption}
\title{From Prompting to Describing: A Cross-Cultural Study of Language for AI-Generated Music}

\twoauthors
   {Sangheon Park} {Georgia Institute of Technology \\School of Music \\ {\tt sangheon@gatech.edu}}
   {Claire Arthur} {Georgia Institute of Technology \\School of Music \\ {\tt claire.arthur@gatech.edu}}

\def\authorname{S. Park, C. Arthur}

\begin{document}

\maketitle

\begin{abstract}

Text-to-music (TTM) generation systems allow users to create music through natural language prompts, yet it is unclear whether the descriptive language used to prompt aligns with descriptive language used to summarize or describe heard music. We pair 200 real-world Udio~\cite{Udio} prompts with their generated audio and free-form descriptions collected from English- ($n = 70$) and Korean-speaking ($n = 78$) listeners, and contribute a human-derived taxonomy of musical prompting vocabulary grounded in real user data. Using this framework, alongside word- and vector-level analyses, we find a consistent structural asymmetry: prompts are dominated by \textit{Genre} and \textit{Story/Narrative} language. Genre terms propagate most reliably from prompt to perception, while narrative-heavy prompts are the strongest predictor of semantic misalignment. A preliminary cross-cultural comparison further suggests that description profiles vary across listener populations along narrative, functional, and affective dimensions, raising questions about whether current TTM systems, trained on aggregated English-centric corpora, can accommodate the full diversity of how people naturally express musical ideas.
\end{abstract}

\section{Introduction}\label{sec:introduction}

Recent text-to-music (TTM) systems such as Suno~\cite{Suno}, Udio~\cite{Udio}, and Google's MusicFX~\cite{GoogleMusicFX} allow users to generate music from a natural language prompt, lowering the barrier to music creation for users without specialized musical knowledge~\cite{electronics14061197}. Understanding how users write these prompts is therefore a fundamental question for music information retrieval and human-AI interaction research. Yet, prompting a generative system is not the same cognitive or linguistic act as describing music one has heard~\cite{mccormack2023}. When a user writes a prompt, they encode anticipatory intent to steer the system toward a desired or imagined output, whereas when a listener describes a piece of music, their language is grounded in a perceptual experience. We argue that these two acts produce systematically different language---a distinction that, despite being intuitive, has not been quantitatively examined. This gap has practical consequences: recent TTM models are trained predominantly on metadata-centric corpora (genre labels, BPM, descriptive tags)~\cite{copet2024simplecontrollablemusicgeneration} rather than on the kind of language users naturally produce when listening, leaving the prompt-description gap unexamined. 
Moreover, this gap limits our ability to evaluate and improve human–AI interaction in music generation, as current systems are optimized for prompt input but commonly assessed through human perception.

A second concern is cultural. Existing TTM datasets are heavily biased toward English and Western-centric conventions~\cite{mehta2025}, yet language is a product of culture---so cross-cultural differences likely shape both how users prompt and how listeners describe. Accordingly, this paper investigates (a) categorical differences between prompting and describing (in English), and (b) cross-cultural differences in description, between native English-speaking and native Korean-speaking listeners. 
We investigate these differences via a listening experiment, in which we presented 200 Udio audio clips, generated from prompts in the Casini et al.~\cite{Casini-2026} dataset, and collected free-form descriptions from both English and Korean participants. 
We investigated two types of similarity: prompt to description similarity (for English participants only, since the prompts were natively in English), and description to description similarity (for English to Korean participants).
We compared the similarity at three levels: a category-level analysis grounded in a human-derived taxonomy, a word-level analysis of lexical propagation and cross-modal associations, and a vector-level analysis of semantic similarity using Sentence-BERT. Our findings reveal systematic differences in how generative intent and perceptual experience are verbalized, with implications for the design of text-conditioned music generation systems and their cross-cultural accessibility. 


\section{Background}
Research on prompt construction has emerged primarily in the text-to-image domain. Oppenlaender~\cite{Oppenlaender17112024} identified six types of prompt \textit{modifiers} through ethnographic study, while subsequent work documented increasing lexical homogenization as user communities converge on effective patterns~\cite{xie2023}. Prompt analysis in text-to-music remains comparatively underdeveloped: Casini et al.~\cite{Casini-2026} provide the primary empirical foundation by collecting real-world TTM prompts and constructing a prompt taxonomy via automated clustering, while Zang and Zhang~\cite{zang2024} conceptualize human-AI music interaction as a problem of ``interpretive alignment.'' Our study builds directly on Casini et al.'s empirical foundation while addressing the prompt-perception comparison that prior work has not examined.

 Several empirical user studies validate the practical impacts in using TTM: PAGURI~\cite{electronics14173379} documented frequent mismatches between musicians' prompting intent and TTM output. 
 For instance, Choi et al.~\cite{choi2025} showed that prompt-based interfaces, while accessible, fail to capture nuanced musical intentions, and Yakura and Goto~\cite{yakura2023} highlighted the ill-defined nature of TTM prompting through their iterative exploration tool. While these studies document the symptom of prompt-output mismatch, they stop short of analyzing the linguistic structures themselves.
Moreover, how people describe music is not culturally uniform. Morrison and Yeh~\cite{morrison1999} found that U.S.\ listeners tend to describe music in structural and analytical terms while Chinese listeners draw more on metaphorical and associative language, and broader work on cross-cultural music cognition~\cite{MORRISON200967} confirms that perceptual attention and descriptive conventions vary across populations. If descriptive language is culturally shaped, then the prompt-description gap may itself vary across listener groups --- a possibility that current TTM research, conducted almost exclusively in English, has not examined. Our study addresses both gaps: characterizing the linguistic structure of the prompt-description mismatch and exploring whether descriptions vary cross-culturally. 

\section{Methods}\label{sec:methods}
\subsection{Taxonomy Construction}
In order to compare descriptions, it was important to have a rubric or taxonomy with which to base the comparison.
While Casini et al. \cite{Casini-2026} did build a taxonomy, theirs was entirely generated automatically by a model.
Given that we wished to ground our comparison in human perception, we sought to create a similar taxonomy, derived from prompting language, but constructed by human annotators.
Accordingly, two doctoral students from Georgia Institute of Technology 
independently analyzed a randomly-sampled subset of 100 prompts from the Casini et al.\ dataset. 
The annotators were instructed to come up with as many labels (or sub-categories) as they deemed necessary to describe all of the prompting language, but that categories needed to be mutually exclusive, and comprehensive. 
The annotators then convened to reconcile differences and iteratively refine the scheme until reaching full consensus on a classification. In the end, this entailed one solo round and two convening rounds to come up with the final taxonomy (the first illustrated ambiguity and non-exclusivity). 
The resulting taxonomy (see Table \ref{tab:taxonomy}) comprises seven categories --- \textit{Genre}, \textit{Mood/Emotion}, \textit{Instrumentation}, \textit{Music Theory}, \textit{Timbre}, \textit{Function}, and \textit{Story/Narrative} --- each defined by explicit inclusion criteria capturing a distinct dimension of musical description.

\begin{table*}[t]
  \centering
  \small
  \begin{tabular}{p{1.8cm}p{4.0cm}p{6.0cm}}
    \hline
    \textbf{Category} & \textbf{Description} & \textbf{Examples} \\
    \hline
    Genre & 
    Labels denoting musical genre, style, or era & 
    \textit{jazz, hip-hop, lo-fi, 80s synthwave, baroque} \\
    \hline
    Mood/Emotion & 
    Affective or emotional qualities of the music & 
    \textit{melancholic, uplifting, tense, dreamy, aggressive} \\
    \hline
    Instrumentation & 
    References to specific instruments or vocal types & 
    \textit{guitar, piano, drum kit, female vocals, strings} \\
    \hline
    Music Theory & 
    Technical or analytical musical attributes such as tempo, key, and structure & 
    \textit{120 bpm, minor key, 4/4 time, chord progression, verse-chorus} \\
    \hline
    Timbre & 
    Perceptual qualities of sound texture and tone color & 
    \textit{heavy, distorted, warm, bright, gritty, airy} \\
    \hline
    Function & 
    Contextual use case or situational role of the music & 
    \textit{background music, workout playlist, film score, lullaby} \\
    \hline
    Story/Narrative & 
    Lyrical content, character, scene, or narrative framing & 
    \textit{love story, heartbreak, rainy night, protagonist, journey} \\
    \hline
  \end{tabular}
  \caption{Human-derived prompt taxonomy. Categories derived via independent dual-annotator coding.}
  \label{tab:taxonomy}
\end{table*}

\subsection{Listening Study}
\subsubsection{Participants}

Data were collected through a human-subjects experiment approved by the Georgia
Institute of Technology Institutional Review Board (IRB 2026-65). English-speaking participants ($n = 71$) were recruited from the Georgia Tech psychology student pool and compensated with course credit (Sona); Korean-speaking participants ($n = 83$) were recruited in two waves, an uncompensated volunteer wave ($n = 33$) and a Prolific wave ($n = 50$). Participants who submitted no descriptions were excluded (one English, five Korean), leaving 148 participants (English $n = 70$; Korean $n = 78$) and $2{,}624$ descriptions (English $= 1{,}396$; Korean $= 1{,}228$) across 200 stimuli.

\subsubsection{Stimuli Selection}
For the stimuli in our listening experiment, we used the Casini et al.~\cite{Casini-2026} corpus of real-world Udio prompts to generate material in Udio, selecting the initial 32-second audio clips.\footnote{Dataset and audio links: \url{https://github.com/mister-magpie/aims_prompts}}
We chose Udio over comparable systems because it maintains a more direct mapping between text prompt and resulting audio; Suno's pipeline, for example, involves extensive LLM-based re-prompting that obscures the original input. We restricted stimuli to the initial 32-second segment, as this is generated directly from the user's prompt prior to any extensions and therefore constitutes the purest acoustic realization of the original query. 
Because the original corpus contains prompts in multiple languages (German, Spanish, Portuguese, Russian), we retained only English-language prompts. 
We manually screened all prompts and embedded lyrics to exclude profanity or sexually explicit content, and discarded clips judged to be of low acoustic quality or non-musical. From the remaining eligible stimuli, 200 clips were randomly selected as the final stimulus set. Because the Casini dataset distributes only prompts and URLs rather than audio, each track was accessed and generated via its original Udio link.\footnote{Audio was collected for non-commercial academic research; we did not redistribute any material. Such use falls within the text and data mining exception of EU Copyright Directive 2019/790, Article 3.}

\subsubsection{Procedure}
The study was conducted online. Participants listened to each clip and described what they heard in free-form natural language. To avoid biasing responses, we showed three example descriptions from the Song Describer dataset~\cite{manco2023song} during a short practice phase, illustrating a range of descriptive styles without prescribing any.
In the experiment proper their instructions were: ``In your own words, describe the music in sufficient detail that the description could be used to distinguish the musical example from another one that may be similar. You must use a minimum of 10 words in your description. You should aim for 10-25 overall.''

English-speaking participants were assigned 20 stimuli as were Korean-speaking participants recruited through Prolific. Voluntary Korean-speaking participants were assigned 10 stimuli to limit burden on uncompensated participants. Demographic information (age, gender, native language, musical experience, and AI familiarity) was collected after the main task.

\begin{figure*}[t]
  \centering
  \includegraphics[width=\textwidth]{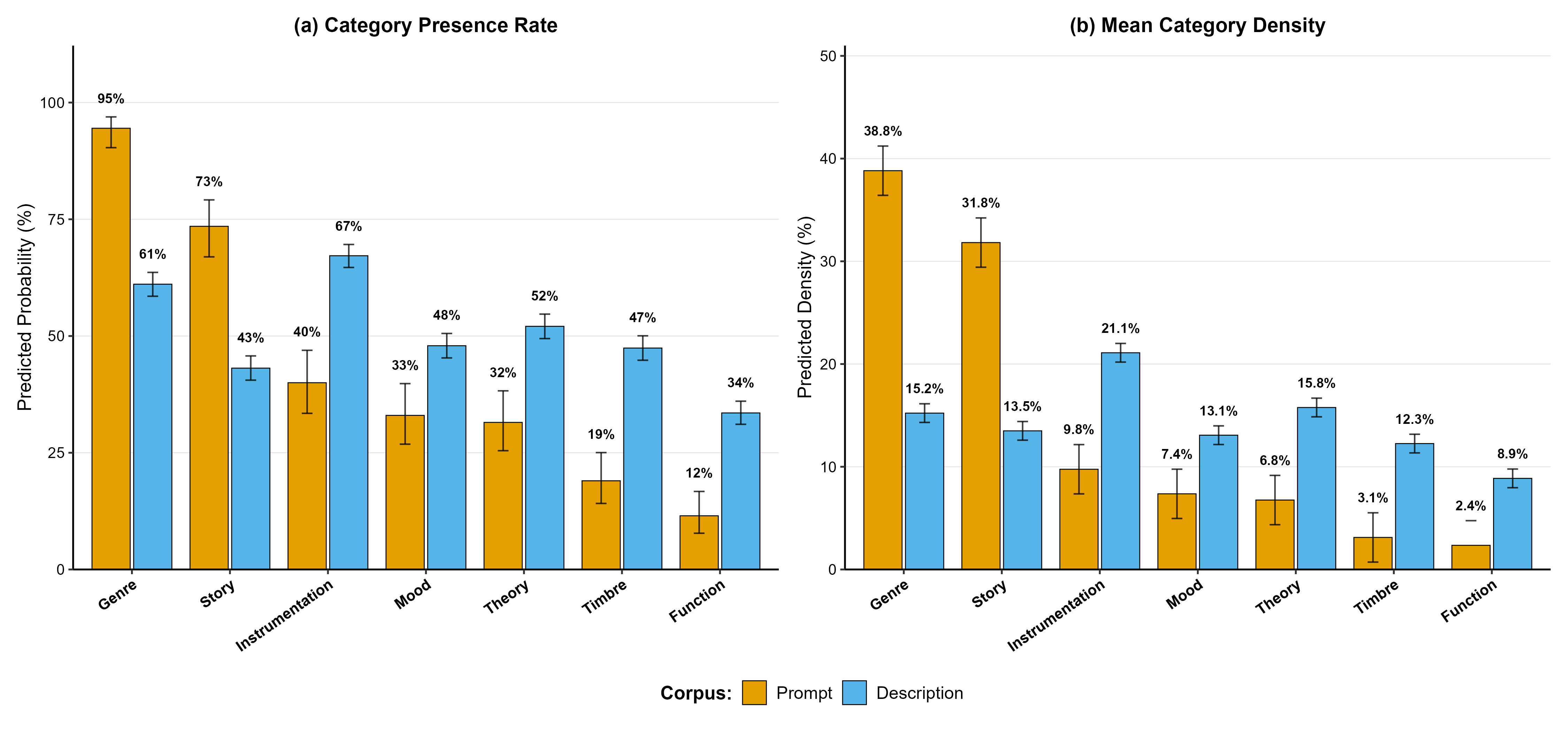}
  \caption{Category-level prompt vs.\ description comparison: 
(a) mean density, (b) presence rate. Prompts dominate in 
\textit{Genre} and \textit{Story/Narrative}; descriptions in 
\textit{Instrumentation}, \textit{Mood/Emotion}, \textit{Music Theory}.}
  \label{fig:category}
\end{figure*}

\section{Results}\label{sec:analysis}

\subsection{Taxonomic Analysis}
To compare prompts and descriptions, we first coded all texts according to our taxonomy. Because manual coding at scale is highly laborious, we employed GPT-5.4. The GPT-5.4 model was selected following a pilot evaluation in which multiple LLM configurations were compared on a held-out, manually-annotated subset 25 of prompts from the Song Describer dataset. This approach is consistent with recent evidence that LLMs can serve as reliable classifiers for text annotation tasks when validated against human judgments~\cite{bunt2025}. To measure categorical differences using our taxonomy, we divided our analysis into two metrics influenced by standard content analysis methodologies~\cite{neuendorf2017, pennebaker2015}: \textit{presence}, to measure the binary selection of essential musical concepts, and \textit{density}, to measure the linguistic focus allocated to each category. 


Because each text contributes seven non-independent observations (one per category), each participant contributes multiple texts which share an audio stimulus, we use mixed-effects regression to account for both sources of non-independence. All models are fit in R using \texttt{lme4}~\cite{JSSv067i01} with random intercepts for text, participant, and stimulus:

\begin{equation}
g(y_{ijkl}) = \beta_0 + \beta_1 C_i + \beta_2 K_j + \beta_3 (C_i \times K_j)
    + u_i + v_k + w_l
\end{equation}



where $g(\cdot)$ is the logit link for presence and the identity link for density (which adds a residual $\varepsilon_{ijkl}$), and $u_i$, $v_k$, $w_l$ are text-, stimulus-, and participant-level random intercepts; prompts have no human author, so $w_l$ is estimated from descriptions only. The Corpus (prompt vs.\ description) $\times$ Category (Taxonomy items) interaction is the quantity of theoretical interest: a significant interaction indicates that the prompt--description difference varies across taxonomic categories. Baseline levels are \textit{description} and \textit{Genre}.

\subsubsection{Presence Analysis}
For each text description (and prompt), we assigned a binary variable,
\textit{presence}, denoting whether at least one word in the text fell
into a given category from the taxonomy. 
We fit this binary input into a GLMM (binomial, logit link) with the
\texttt{bobyqa} optimizer, using the fixed- and random-effects structure
specified above.
The fitted model highlights asymmetries between corpora. In the descriptions, \textit{Genre} appears in roughly three-fifths ($\hat{p} = 0.615$),
with \textit{Instrumentation} the only category significantly more common than Genre ($\beta = 0.28$, $p < .001$); all other categories are significantly rarer than genre. In the prompts, \textit{Genre} becomes near-ubiquitous ($\hat{p} \approx 0.95$), a substantially larger presence than in the descriptions, and the clearest difference in prompt convention.


Because the model includes an interaction term, the category coefficients are read relative to 
Genre: each one indicates how much a category's prompt--description gap differs from the gap observed for Genre. Among all categories, \textit{Story/Narrative/Lyrics} has the least negative coefficient after Genre ($\beta = -1.07$, $z = -2.98$, $p = .002$). In other words, \textit{Story/Narrative} is the one category, like Genre, that tends to appear more in prompts than descriptions ($\hat{p} \approx 0.74$).  The remaining five categories show strongly negative interactions ($\beta$ ranging $-3.00$ to $-3.72$, all $p < .001$): \textit{Instrumentation}, \textit{Mood/Emotion}, \textit{Music Theory}, \textit{Timbre}, and \textit{Function} are not merely less elevated in prompts but fall below their description rates. Consistent with Lesaffre et al.~\cite{lesaffre2006}, this suggests that the perceptual and affective dimensions central to listening are systematically reduced when users write generative prompts.

Recall, however, that \textit{presence} captures which categories are used, but not how heavily they
dominate a text. We therefore turn to \textit{density}.

\subsubsection{Density Analysis}
Category density is the proportion of a text's words belonging to a given taxonomic category $C$ (the category's word count divided by the text's total word count). While presence measures whether a category is invoked, density approximates how much lexical space it occupies. We fit an LMM with the same fixed- and random-effects structure as the presence model.

The density results reinforce and refine the presence findings. In descriptions, \textit{Genre} accounts for 15.2\% of vocabulary, and \textit{Instrumentation} 21.1\%. In prompts, the genre share rises to 38.8\% ($\beta = 0.236$, $t = 18.03$, $p < .001$), roughly two and a half times the description rate. Interactions again single out \textit{Story/Narrative}: it preserves most of the genre-level prompt elevation ($\beta = -0.053$, $t = -2.85$, $p = .004$, about one-sixth the magnitude of the other interactions), while the remaining five categories show near-complete reversal ($\beta$ from $-0.29$ to $-0.35$, all $p < .001$). Perceptual and affective categories are systematically denser in descriptions than in prompts.

\subsection{Word-Level Analysis}
At the word level, we removed stop words (i.e., articles, conjunctions, etc.), and lemmatized the remaining tokens before analysis. 


We first computed a per-term survival rate: for each prompt word, the proportion of its songs whose paired descriptions reused that word. Genre terms spanned a broad range (\textit{country}: 73.8\%, \textit{rock}: 36.2\%), overlapping with concrete instrument terms (\textit{guitar}: 37.6\%), while affective and narrative terms (\textit{upbeat}, \textit{love}) propagated poorly. To examine whether specific prompt vocabulary systematically elicits particular description vocabulary, we partitioned songs by whether their prompt contained a target word $w_p$ and tested for each description word $w_d \neq w_p$, whether it occurred more often in the $w_p$-present group via Pearson's chi-squared test. To keep expected cell counts adequate, we retained only pairs in which $w_p$ occurred more than ten times across the prompt corpus and was paired with at least twenty descriptions, and in which $w_d$ co-occurred with $w_p$ at least three times; only positive associations are reported.

Results (Table~\ref{tab:chi2}) reveal a striking asymmetry in what survives the prompt--audio--description chain. Every prompt word meeting our frequency threshold belongs to \textit{Genre} or \textit{Instrumentation}; no \textit{Mood/Emotion} or \textit{Story} term produced a reliable association, consistent with the subjectivity and lexical diversity of affective response. Even for surviving terms, listeners rarely reused the prompt word itself, substituting an adjacent label instead (\textit{metal} $\to$ \textit{rock}, \textit{punk}; \textit{hip-hop} $\to$ \textit{rap}; \textit{house} $\to$ \textit{techno}, \textit{edm}; \textit{country} $\to$ \textit{folk}) or an associated sonic marker (\textit{rock} $\to$ \textit{guitar}; \textit{metal} $\to$ \textit{electric}, \textit{scream}; \textit{country} $\to$ \textit{banjo}). Prompt language is thus recovered as a semantic neighborhood rather than as specific words.


\begin{table}[t]
  \centering
  \small
  \begin{tabular}{llc}
    \hline
    \textbf{Prompt} & \textbf{Description} & \textbf{$\chi^2$} \\
    \hline
    heavy       & metal     & 191.66 \\
    hip-hop     & rap       & 184.04 \\
    metal       & rock      & 134.59 \\
    dance       & jackson   & 103.22 \\
    funk        & jackson   & 101.82 \\
    blue        & lullaby   &  99.69 \\
    country     & southern  &  86.99 \\
    r\&b        & duet      &  79.78 \\
    soul        & duet      &  76.70 \\
    piano       & lullaby   &  72.70 \\
    male        & rock      &  65.79 \\
    bass        & spanish   &  64.74 \\
    rock        & guitar    &  63.55 \\
    jazz        & french    &  58.72 \\
    \hline
  \end{tabular}
  \caption{Strongest cross-modal prompt--description word associations.}
  \label{tab:chi2}
\end{table}

Beyond categorical content, we asked whether prompts and descriptions 
differ in the \emph{concreteness} of their vocabulary. Using the 40,000-word concreteness norms of Brysbaert et al.~\cite{brysbaert2014concreteness}, where ratings range from 1 (highly abstract) to 5 (highly concrete), we computed the mean concreteness of all rated content words in each  text.\footnote{80.3\% of prompt tokens and 91.4\% of description tokens matched the Brysbaert lexicon; unmatched tokens (e.g., genre neologisms, proper nouns) were excluded from per-text means.}

Prompts ($M = 3.52$, $SD = 0.44$) were significantly more concrete 
than their paired descriptions ($M = 3.44$, $SD = 0.19$; Wilcoxon $W = 7907$, $p = .009$, paired by song; $n = 200$). This register-level asymmetry aligns with and extends the category findings: prompts rely on concrete acoustic anchors (genre labels, instruments) while descriptions draw more heavily on abstract, affective, and imagistic vocabulary --- the linguistic register through which listeners naturally express perceptual experience. Notably, neither prompt concreteness ($r = -0.014$, $p = .84$) nor description concreteness ($r = -0.101$, $p = .15$) correlated with prompt-description cosine similarity, suggesting that concreteness alone does not predict semantic alignment --- consistent with our narrative finding that even highly concrete narrative prompts (scenes, characters) can fail to produce acoustically recoverable output.

\subsection{Vector-Level Analysis}

We encoded prompts and descriptions into a shared embedding space 
using Sentence-BERT~\cite{reimers2019sentence} and computed, for each prompt, the mean cosine similarity to all of its paired descriptions as a measure of semantic alignment. To identify which descriptions were closest to prompts (and which had the widest gap), we compared the top 25\% (high-alignment, $n = 50$) against the bottom 25\% (low-alignment, $n = 50$) on category presence and density.

Results are summarized in Table~\ref{tab:vector}. Despite \textit{Genre} being the dominant category in both groups (high: 100\%, low: 84\%), its density is nearly identical between high- and low-alignment groups (0.373 vs.\ 0.357), suggesting that genre labeling alone neither closes nor widens the semantic gap. By contrast, \textit{Mood/Emotion}, \textit{Instrumentation}, and \textit{Music Theory} appear roughly two to four times more often in the high-alignment group (presence: $\Delta = +0.36$, $+0.52$, and $+0.22$, respectively), indicating that prompts specifying affective character, sonic content, and structural attributes are more likely to produce audio that listeners describe in semantically congruent terms.

\begin{table}[t]
  \centering
  \small
  \begin{tabular}{lcccc}
    \hline
    & \multicolumn{2}{c}{\textbf{High 25\%}} 
    & \multicolumn{2}{c}{\textbf{Low 25\%}} \\
    \textbf{Category} & Pct. & Den. & Pct. & Den. \\
    \hline
    Genre               & 100.0 & 0.373 & 84.0 & 0.357 \\
    Instrumentation     &  70.0 & 0.169 & 18.0 & 0.050 \\
    Story/Narrative     &  60.0 & 0.169 & 80.0 & 0.452 \\
    Mood/Emotion        &  52.0 & 0.113 & 16.0 & 0.042 \\
    Music Theory        &  48.0 & 0.098 & 26.0 & 0.067 \\
    Timbre              &  38.0 & 0.060 &  6.0 & 0.014 \\
    Function            &  14.0 & 0.020 &  6.0 & 0.019 \\
    \hline
  \end{tabular}
  \caption{Category presence rates (\%) and mean densities for high- and 
low-alignment prompt groups (top and bottom 25\% by mean 
prompt--description cosine similarity), sorted by high-alignment 
presence.}
  \label{tab:vector}
\end{table}

The most striking divergence concerns \textit{Story/Narrative}. Although it appears in 80\% of low-alignment prompts (more frequent than in high-alignment ones), its density is nearly three times higher in the low-alignment group (0.452 vs.\ 0.169). This pattern has a structural explanation: narrative language encodes intent (a scene, a character, a story) that leaves no recoverable acoustic trace, so narrative-heavy prompts produce audio that listeners describe in semantically misaligned terms --- not because generation failed, but because the intent was never acoustically encodable. Listeners hearing only the audio describe what they perceive, not what the prompt imagined.

\begin{figure*}[t]
  \centering
  \includegraphics[width=\textwidth]{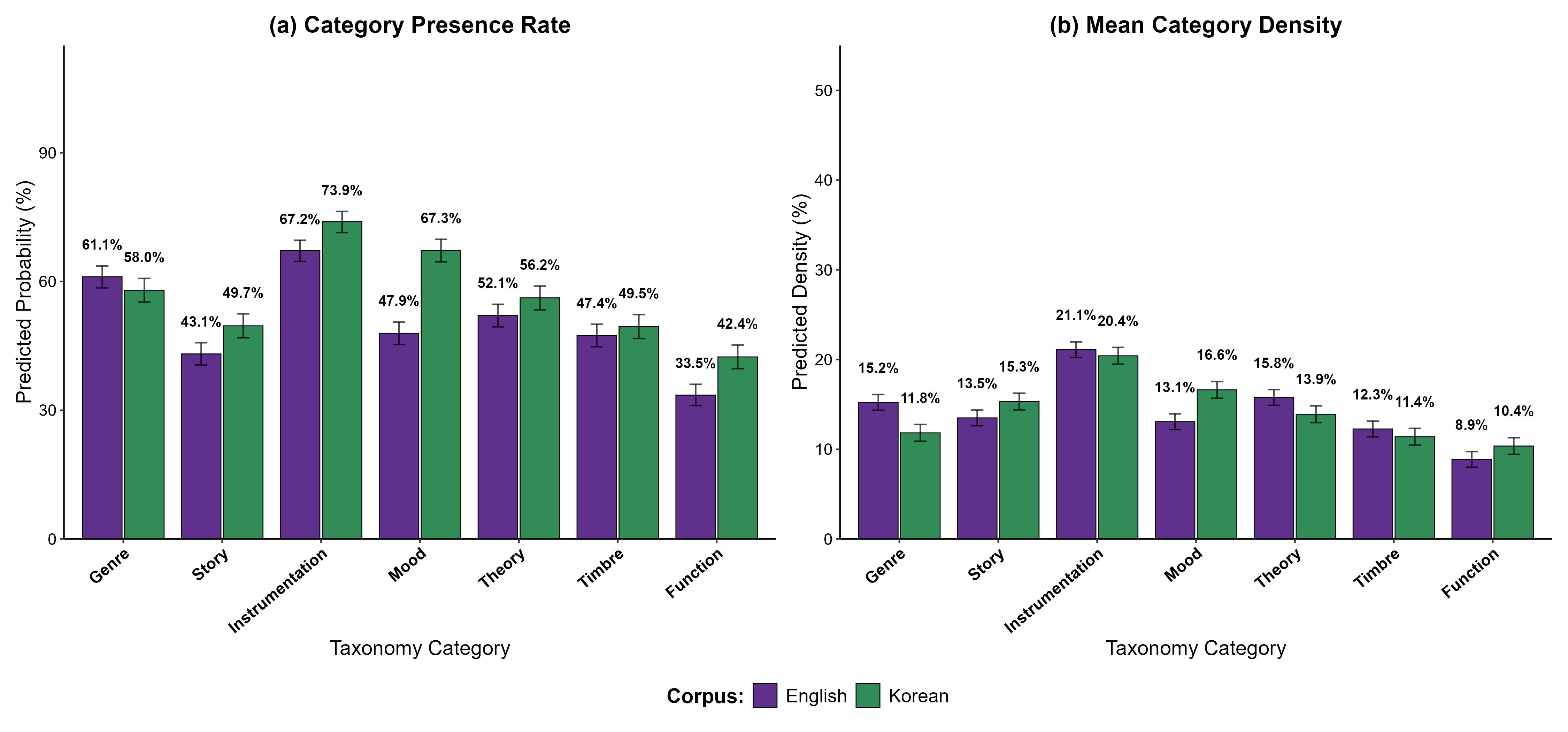}
  \caption{Cross-cultural description comparison by language group
(purple: English, $n = 70$; green: Korean, $n = 78$):
(a) model-predicted category presence rate, (b) mean category density.
Error bars show 95\% confidence intervals.}
  \label{fig:kr_en}
\end{figure*}

\subsection{Cross-Cultural Analysis}

To explore whether description profiles vary across listener populations, we compared the descriptions from the English-speaking ($n = 70$) and Korean-speaking ($n = 78$) participants. Because the two groups differ in recruitment methods, and we did not have any specific hypotheses regarding group differences, we treat these results as exploratory. 

We refit the same GLMM and LMM specifications used above, replacing \textit{corpus} with \textit{language group} (English vs.\ Korean). At the genre level baseline, Korean descriptions showed slightly lower density than English ($\beta = -0.034$, $t = -5.21$, $p < .001$), while the presence difference was not significant ($\beta = -0.155$, $z = -1.49$, $p = .136$). 

The language $\times$ category interactions were uniformly positive: relative to the genre baseline, every category was more elevated in Korean than in English. 
Because this pattern is broad rather than selective, we report simple effects within each category. In \textit{presence}, Korean descriptions invoked more of the taxonomy in four categories --- \textit{Mood/Emotion} ($\mathit{OR} = 0.45$, $p < .001$), \textit{Function} ($\mathit{OR} = 0.68$, $p < .001$), \textit{Instrumentation} ($\mathit{OR} = 0.73$, $p = .004$), \textit{Story/Narrative} ($\mathit{OR} = 0.77$, $p = .013$); \textit{Genre} ($p = .136$), \textit{Music Theory} ($p = .139$), and \textit{Timbre} ($p = .503$) showed no difference.

In \textit{density}, the differences resolve into a systematic redistribution
rather than a uniform shift. English descriptions allocated more lexical space to \textit{Genre} (15.2\% vs.\ 11.8\%; $p < .001$) and \textit{Music Theory} (15.8\% vs.\ 13.9\%; $p = .004$), whereas Korean descriptions allocated more to \textit{Mood/Emotion} (16.6\% vs.\ 13.1\%; $p < .001$), \textit{Story/Narrative} (15.3\% vs.\ 13.5\%; $p = .005$), and \textit{Function} (10.4\% vs.\ 8.9\%; $p = .023$). \textit{Instrumentation} ($p = .295$) and \textit{Timbre} ($p = .188$) did not differ.


These findings suggest that cultural and linguistic background may shape both the breadth and the emphasis of musical description. Korean listeners invoked more dimensions of the taxonomy overall, and redistributed lexical space along a consistent axis: away from the categorical vocabulary of genre labels and structural attributes, and toward affective, narrative, and functional framing. This is consonant with Morrison and Yeh's~\cite{morrison1999} observation that U.S.\ listeners describe music in structural terms while East Asian listeners draw more on metaphorical and associative language.



\section{Discussion}

\subsection{The Prompt-Description Gap}


Prompts and descriptions are not two registers of the same language but structurally different communicative acts. Prompts are dominated by \textit{Genre}
and \textit{Story/Narrative/Lyrics} (95\% and 74\% presence respectively),
whereas descriptions are richer in \textit{Instrumentation}, \textit{Mood/Emotion}, and \textit{Music Theory} --- the perceptual and affective dimensions listeners actually attend to when experiencing music.

The word- and vector-level analyses identify a narrow but consistent subset of vocabulary that bridges generative intent and perceptual experience. Genre terms reliably survive the chain from text to audio to perception, though often as a semantic neighborhood rather than the label itself --- realized acoustically through constituent sonic markers (e.g., \textit{rock} $\to$ \textit{guitar}, \textit{drum}) which listeners then name independently.


\textit{Story/Narrative} vocabulary, by contrast, exhibits the weakest semantic alignment, with density nearly triple in low-alignment prompts (0.452 vs.\ 0.169). Prior work has shown that listeners with shared cultural backgrounds produce remarkably consistent narrative imaginings in response to instrumental music~\cite{margulis2022, MCAULEY2021104712}, so the failure of such cues to survive from prompt to description suggests that current TTM systems do not encode the acoustic features supporting shared narrative perception. While some prompts contain humorous or imagery-based content that even skilled composers would find difficult to realize sonically, we view their presence as further evidence of the 
structural argument: when prompts encode intent without an acoustic correlate, the resulting audio cannot recover that intent regardless of generative capacity. 

\subsection{One system, Many Listeners}

Beyond the prompt-description gap, our results reinforce that descriptions of the same audio vary substantially across listeners, shaped by perceptual attention, musical background, and preliminarily cultural framing~\cite{manco2023song, 10.1371/journal.pone.0089642, morrison1999}. Current TTM systems flatten this variation into a single training register, so users whose natural vocabulary draws on narrative, affective, or functional terms face a higher --- linguistic and cultural --- threshold for being understood. Our cross-linguistic comparison suggests this variation is patterned rather than random: Korean listeners systematically reallocated lexical space from genre and structural vocabulary toward affective and narrative framing.



\subsection{Limitations and Future Work}

Several limitations bound our findings. Although the two language groups are comparable in size ($n = 70$ and $n = 78$), they differ in recruitment channel: English-speaking participants were drawn from a single university population, while Korean participants were recruited through both personal networks and Prolific. The cross-cultural findings may therefore reflect differences in participant pool composition such as age or musical training, as much as language or culture.



All stimuli were generated by a single TTM system (Udio), leaving open whether the gap we characterize generalizes across systems or partly reflects Udio-specific generation behavior. Finally, our automated LLM-based labeling, though validated against human judgments, introduces classification noise whose downstream effects on the mixed-effects models are difficult to fully quantify.

\section{Conclusion}
We characterized the linguistic gap between prompting text-to-music systems and describing the resulting audio. Prompts and descriptions emerge as structurally distinct registers: \textit{Genre} and \textit{Story/Narrative} dominate prompts, while \textit{Instrumentation}, \textit{Mood/Emotion}, and \textit{Music Theory} dominate descriptions. Genre vocabulary propagates most reliably from prompt to perception, while narrative-heavy prompts are the strongest predictor of semantic misalignment. The promise of TTM as an accessible creative technology therefore depends on whether systems can accommodate the full diversity of musical expression.

\section{AI Usage Statement}
Large language models were used as a component of the study's methodology. As described in Section~\ref{sec:analysis}, GPT-5.4 was used to assign taxonomic category labels to all prompt and description texts; model selection, prompting procedure, and validation against human-coded data are reported there and in the supplementary material. The taxonomy itself was constructed entirely by human annotators. LLMs were additionally used for editorial assistance in drafting and revising the manuscript; no figures, tables, or references were AI-generated. All claims, interpretations, and citations were verified by the authors, who take full responsibility for the correctness, originality, and integrity of the manuscript.

\bibliography{ISMIRtemplate}
\end{document}